\documentclass[aps,prd,twocolumn,amsmath,amssymb,superscriptaddress]{revtex4-1}

\usepackage{graphicx}
\begin{document}

\title{Equation of state for the MCFL phase and its implications for compact star models}
\author{L. Paulucci}
\email{laura.paulucci@ufabc.edu.br}
\affiliation{Universidade Federal do ABC \\
Rua Santa Ad\'elia, 166, 09210-170 Santo Andr\'e, SP, Brazil}
\author{Efrain J. Ferrer}
\email{ejferrer@utep.edu}
\author{Vivian de la Incera}
\email{vincera@utep.edu}
\affiliation{Department of Physics, University of Texas at El Paso,
El Paso, TX 79968, USA}
\author{J. E. Horvath}
\email{foton@astro.iag.usp.br}
\affiliation{Instituto de Astronomia, Geof\'\i sica e Ci\^encias
Atmosf\'ericas\\ Rua do
Mat\~ao 1226, 05508-900 S\~ao Paulo SP, Brazil}

\date{\today}

\begin{abstract}
Using the solutions of the gap equations of the {magnetic-color-flavor-locked} (MCFL) phase of paired quark matter
in a magnetic field, and taking into consideration the separation between the longitudinal and transverse
pressures due to the field-induced breaking of the spatial rotational symmetry,
the equation of state (EoS) of the MCFL phase is self-consistently
determined. This result is then used to investigate the possibility of absolute stability,
which turns out to require a field-dependent ``bag constant'' to hold. That is,
only if the bag constant varies with the magnetic field, there exists a window in the magnetic field vs. bag
constant plane for absolute stability of strange matter. Implications
for stellar models of magnetized (self-bound) strange stars and
hybrid (MCFL core) stars are calculated and discussed.
\end{abstract}


\maketitle

\section{Introduction}

Neutron stars are dense, compact astrophysical objects which are
one possible result of the evolution of massive stellar progenitors.
Determining which is the state of the matter in the interior of
these objects is still an open question, and of the greatest
importance for hadronic physics and stellar astrophysics alike. High-quality data
presently being taken and analyzed offer for the first time a real
perspective to explore this domain of strong interactions.

It has been proposed that these stars are not composed of neutron
matter, but rather that, given the conditions of very high density
in their interiors, there could be a phase transition from nuclear
to quark matter \cite{AFO, SS}.
Several authors have considered an even more extreme
possibility \cite{Dinos, Witten}: the absolute
stability of the deconfined phase (in which case, self-bound
-strange stars- would exist). If the milder condition is realized,
that is, the deconfined phase is stable only at high pressure,
stars with quark cores (hybrid stars) would ensue.

An interesting twist to the stability problem was given a decade
ago (after an important precursor \cite{Bailin}),
when paired matter was studied
\cite{Pairing, RajWilc, German} and the pairing
energy was shown to enlarge the window of stability in parameter
space. The phenomenon of color superconductivity, in which
quarks pair according to their color and flavor in a
specific pattern, would thus introduce a pairing gap in the
free energy of the system due to the attractive color-antisymmetric
channel in the interaction between quarks.

The most symmetric pairing state would be the Color-Flavor-Locked
one (CFL) when quarks of all flavors and colors pair. This state
can only be realized if the mass split between the lightest quarks
(up and down) and the strange quark is small and/or the chemical
potential $\mu$ is high enough, a condition usually written as
$\mu\gtrsim m_s^2 / 2\Delta$ \cite{Alford04}, with $m_s$ being the
strange quark mass and $\Delta$ the pairing gap. When this
condition does not hold, other states could be realized
(e. g., LOFF \cite{LOFF}, kaon-condensate phase
\cite{Schafer}, 1SC \cite{Alford04, 1SC}, homogenous
gluon condensate phase \cite{Gorbar},
and gluon-vortex lattice \cite{vortex}, among others). This is a
subject under intense study.

Being a possible physical realization of dense matter physics,
a common characteristic of neutron stars is their strong
magnetization. Their surface magnetic fields range from $H =
1.7\times10^8G$ (PSR B1957+20) up to $2.1\times10^{13}G$ (PSR
B0154+61), with a typical value of $10^{12}G$ \cite{NS-H}. There
are observational evidences of even stronger magnetic fields in
the special group of neutron stars know as magnetars- with
surface magnetic fields of order $B \sim 10^{14}-10^{15}G$
\cite{Magnetar}. In the core of these compact
objects the field may be considerably larger due to flux
conservation during the core collapse or by internal mechanisms that
can boost a pre-existing seed field
\cite{Int-Mech}. By applying the equipartition theorem, the
interior field can be estimated to reach values $H \sim 10^{19-20}G$
\cite{H-Estimate}. Therefore, if color superconducting QCD phases
constitute the neutron matter interiors, it is likely that a
treatment including high field values would be needed.

At this point, it is worth to underline a main difference
between a conventional electric
superconductor and a spin-zero color superconductor in regard to
their behavior in the presence of a magnetic field. Spin-zero
color superconductivity, as that of the CFL phase and the
two-flavor 2SC phase, does not screen an external magnetic field
because even though the color condensate has non-zero electric
charge, there is a linear combination of the photon and the 8th
gluon,
$\widetilde{A}_{\mu}=A_{\mu}\cos\theta-G_{\mu}^8\sin\theta$, that
remains massless \cite{Gatto}. This combination plays the role of
an in-medium or rotated electromagnetism with the color condensate
being always neutral with respect to the corresponding rotated
charge $\widetilde{Q}$. Then, an external magnetic field can
penetrate the color superconductor through its long-range,
in-medium component $\widetilde{H}$. Furthermore, even though the
diquark condensate is neutral with respect to the rotated
electromagnetism, some quarks participating in the pairing are
$\widetilde{Q}$-charged, so they can couple to a background
magnetic field thereby affecting the gap equations of the system
\cite{Cristina-1, Cristina-2, Cristina-3, Sadooghi}. Because of this effect, the
three-flavor color superconductor in a magnetic field exhibits a
new phase that is known as Magnetic Color-Flavor-Locked (MCFL)
phase \cite{Cristina-1}. Although the CFL and the MCFL phases of
three-flavor paired quark matter are similar in that they both
break chiral symmetry through the locking of color and flavor and
have no Meissner effect for an in-medium magnetic field, they have
important differences too (for physical implications of their
differences see \cite{Phases}).

At present, some of the best-known characteristics of stellar
objects are their masses and radii. The relation between the mass and the
radius of a star is determined by the equation of state (EoS) of the
microscopic matter phase in the star. If one can find some features that can connect
the star's internal state (nuclear,
strange, color superconducting, etc.) to its mass/radius relation, one
would have an observational tool to discriminate among the actual
realization of different star inner phases in nature. From previous
theoretical studies \cite{German, Alford03, MassRadius}
the mass-radius relationship predicted for neutron stars with
different quark-matter phases (CS or unpaired) at the core are very
similar to those having hadronic phases, at least for the observed mass/radius range.
As a consequence, it
is very difficult to find a clear observational signature that
can distinguish among them. Nevertheless, an important ingredient
was ignored in these studies: the magnetic field, which in some compact
stars could reach very high values in the inner regions.

As pointed out in \cite{H-Estimate},
a strong magnetic field can create a significant
anisotropy in the
longitudinal and transverse pressures.
One would expect then, that the EoS, and consequently,
the mass-radius ratio, become affected by sufficiently strong core fields. Given
that we are beginning to obtain real observational constraints on the
EoS of neutron stars \cite{Alford2010},
it is important to investigate the EoS in the presence of a magnetic
field for different inner star phases to be able to discard those
that do not agree with observations.

In order to understand the relevance of the magnetic field to
tell apart neutron stars from stars with paired quark matter,
it is convenient to recall that when the pressure
exerted by the central matter density of neutron stars (which is about
$200-600 MeV/fm^3$) is contrasted with that exerted by an electromagnetic
field, the field strength needed for these two contributions to be
of comparable order results of order $\sim 10^{18} G$ \cite{Broderick}.
On the other hand, relevant relativistic quantum field effects (i.e.
those due to the Landau quantization of the particle energy modes)
will show up in the neutron/proton star matter when the particles's cyclotron energy
$ehH/mc$ becomes comparable to its rest energy $mc^{2}$,
which for protons means a field $\sim 10^{20} G$.

However, for stars with paired quark matter, the situation
is rather different. Naively, one might think that comparable matter and field pressures
in this case would occur only at
much larger fields, since the quark matter can only exist at even larger densities
to ensure deconfinement. In reality, though, the situation is more subtle. As argued
in \cite{Alford03}, the leading term in the matter pressure coming from the contribution
of the particles in the Fermi sea, $\sim\mu^{4}$, could be (almost) canceled out by the negative
pressure of the bag constant and in such a case, the next-to-leading term would play a
more relevant role than initially expected. Consequently, the magnetic pressure might
only need to be of the order of that produced by the particles close to the Fermi surface,
which becomes the next-to-leading contribution, $\sim\mu^{2}\Delta^{2}$, with $\Delta$ the
superconducting gap and $\mu$ the baryonic chemical
potential. For typical values of these parameters in paired quark matter one obtains a
field strength $\sim 10^{18} G$. Moreover, the magnetic field can affect the pressure
in a less obvious way too, since as shown in Refs. \cite{Cristina-1, Cristina-2, Cristina-3},
it modifies the structure and magnitude of the superconductor's gap, an effect that, as found in \cite{Paramagnetism}, starts to become relevant already at fields of order $10^{17} G$
and leads to de Haas van-Alphen oscillations of the gap magnitude \cite{Noronha, Fukushima}.
It is therefore quite plausible that the effects of moderately
strong magnetic fields in the EoS of compact
stars with color superconducting matter will be more noticeable than in
stars made up only of nucleons, where quantum effects starts to be
significant for field three orders of magnitude larger. This is why an evaluation of a magnetized
quark phase is in order.

In this work, we perform a self-consistent analysis of the EoS of the MCFL matter, taking into consideration the solution of the gap equations and the anisotropy of the pressures in a magnetic field. Our main goals are: 1) to investigate the effect of the magnetic field in the absolute stability of strange stars made of paired matter in the MCFL phase; 2) to determine the threshold field at which substantial separation between the parallel and transverse pressures occurs in the MCFL matter; and 3) to explore whether there is a range of magnetic field  strengths, within the isotropic regime for the EoS, that can lead to observable differences in the mass-radio ratios of stars with MCFL vs CFL cores.

The plan of the paper is the following. In Sec. II we present the
thermodynamic potentials for the color superconducting (CFL and
MCFL) models used in our calculations throughout. Using them, the
equations of state for the CFL and MCFL phases are then found in
Sec. III. The pressure anisotropy appearing in the MCFL case is
graphically shown and the order of the field strength required for
the anisotropic regime to settle is determined. The stability conditions for
the realization of self-bound MCFL matter is investigated in Sec.
IV, where we find that the magnetic field acts as a
destabilizing factor for the realization of strange matter and prove that only
if the bag constant decreases with the field, a magnetized strange
star could exist. In Sec. V, we applied the EoS of MCFL matter to
calculate the mass-radius relationship of self-bound and
gravitational-bound stellar models.
The main outcomes of the paper are summarized in Sec. VI. Finally, in Appendix A, it is studied the dynamical bag constant in the chiral limit at $H\neq 0$.

\section{Model}
As mentioned in the Introduction, a main goal of this work is to
carry out a self-consistent investigation of the EoS of the MCFL phase.
For the sake of understanding, and for comparison with the case
without magnetic field, we are also going to find the EoS of the
CFL phase using a similar approach. With this aim in mind, we
first need to obtain the thermodynamic potential for each
phase. The CFL superconductor can be modeled by the three-flavor
Nambu-Jona-Lasinio (NJL) theory considered in \cite{alf-Kouvaris-raj} (see Eq.(10) of
that reference). In our case, we neglect all the quark masses so
the color and electrical neutralities are automatically satisfied
and the only nonzero chemical potential will be the baryonic
chemical potential $\mu$. As known, this effective model displays all the symmetries of QCD which are relevant at high densities.
Its four-fermion point interaction
contains the quark-quark attractive color antitriplet channel that
gives rise to the diquark condensate.

In the MCFL phase we assume a uniform and constant magnetic field. The reliability of this assumption for neutron stars, where the magnetic field strength is expected to vary from the core to the surface in several orders, is based on the fact that the scale of the field variation in the stellar medium is much larger than the microscopic magnetic scale for both weak and strong magnetic fields \cite{Broderick}. Hence, when investigating the field effects in the EoS, it is consistent to take a magnetic field that is locally constant and uniform. This is the reason why such an approximation has been systematically used in all the previous works on magnetized nuclear \cite{Broderick, MNS} and quark matter \cite{MQS}.

\subsection{Thermodynamic Potential of the CFL Phase}

The mean-field thermodynamic potential of the CFL phase is \cite{alf-Kouvaris-raj}

\begin{equation}\label{omeganofield}
\Omega=-\frac{T}{2}\sum_{n}\int\frac{d^{3}p}{(2\pi)^{3}}Tr\log(\frac{1}{T}S^{-1}(iw_{n},p))
+\frac{\Delta_{\eta}\Delta_{\eta}}{G}
\end{equation}
where the sum in $n$ indicates the finite temperature sum in the
Matsubara frequencies. The inverse full propagator here is
\begin{widetext}
\begin{equation}\label{CFLpropagator}
S^{-1}(p)=\left(
            \begin{array}{cc}
              p\llap/+\mu\llap/ & P_{\eta}\Delta_{\eta} \\
              \overline{P}_{\eta}\Delta^{*}_{\eta} & p\llap/-\mu\llap/ \\
            \end{array}
          \right)
\end{equation}
\end{widetext}
with
$(P_{\eta})_{ij}^{ab}=C\gamma_{5}\epsilon^{ab\eta}\epsilon_{ij\eta}$
(no sum over $\eta$) and
$\overline{P}_{\eta}=\gamma_{4}P^{\dagger}_{\eta}\gamma_{4}$. The
gap is
$\Delta_{\eta}=<\frac{G}{2}\psi^{T}\overline{P}_{\eta}\psi>$ with
quark field $\psi$ of colors ($r,g,b)$ and flavors $(u,d,s)$. The
index $\eta=1,2,3$ labels the d-s, u-s, and u-d pairing
respectively.

After summing in $n$ and taking the zero temperature limit, one obtains
\begin{widetext}
\begin{equation}
\Omega_{CFL} =-\frac{1}{4\pi^2}\int_0^\infty dp p^2 e^{-p^2/\Lambda^2}[16|
\varepsilon|+16|\overline{\varepsilon}|]-\frac{1}{4\pi^2}\int_0^\infty
dp p^2 e^{-p^2/\Lambda^2}[2|\varepsilon'|+2|\overline{\varepsilon'}|]+
\frac{3\Delta^2}{G} \label{OmegaCFL}
\end{equation}
\end{widetext}
where

\begin{eqnarray*}\label{6}
\varepsilon=\pm \sqrt{(p-\mu)^2+\Delta_{CFL}^2} \nonumber \\
\overline{\varepsilon}=\pm \sqrt{(p+\mu)^2+\Delta_{CFL}^2}\nonumber
\\
\varepsilon'=\pm \sqrt{(p-\mu)^2+4\Delta_{CFL}^2}\\
 \overline{\varepsilon'}=\pm \sqrt{(p+\mu)^2+4\Delta_{CFL}^2}\nonumber
\end{eqnarray*}
are the dispersion relations of the quasiparticles. Here we
already took into account the well-known solution
$\Delta_{CFL}=\Delta_{1}=\Delta_{2}=\Delta_{3}$, valid for the CFL
gap at zero quark masses. As in \cite{Noronha}, in order to have
only continuous thermodynamical quantities, we
introduced in (\ref{OmegaCFL}) a smooth cutoff depending on $\Lambda$.

\subsection{Thermodynamic Potential of the MCFL Phase}

Let us consider now the case with a rotated magnetic field $\widetilde{H}$,
which couples to the charged quarks
through the covariant derivative of the NJL Lagrangian. The magnetic interaction leads
to the separation of the original $(u,d,s)$ quark
representation into neutral, positively and negatively
charged spinors according to the quark rotated charges in units of $\widetilde{e}=e\cos\theta$, with $\theta$ being the mixing angle of the rotated fields,
\begin{equation}
\label{q-charges}
\begin{tabular}{|c|c|c|c|c|c|c|c|c|}
  \hline
  $u_{r}$ & $u_{g}$ & $u_{b}$ & $d_{r}$ & $d_{g}$ & $d_{b}$ & $s_{r}$ & $s_{g}$ & $s_{b}$ \\
  \hline
  0 & 1 & 1 & -1 & 0 & 0 & -1 & 0 & 0 \\
  \hline
\end{tabular}
\end{equation}
\\

Because of this separation, it is convenient to introduce three sets of
Nambu-Gorkov spinors that correspond to positive-, negative- and
zero-charged fields. The details of this procedure, as well as a
discussion of Ritus' method \cite{Ritus}, used to transform the charged
spinor fields to momentum space in the presence of a magnetic field, can be found in
\cite{Cristina-2}. After integrating in the fermion fields, doing the
Matsubara sum and taking the zero temperature limit, we can write
the MCFL thermodynamic potential as the sum of the contributions
coming from charged ($\Omega_C$) and neutral ($\Omega_N$) quarks.
\begin{equation}
\Omega_{MCFL} =\Omega_C+\Omega_N
\label{OmegaMCFL}
\end{equation}
\\
with
\begin{widetext}
\begin{equation}\label{C}
\Omega_{C} =-\frac{\widetilde{e}\widetilde{H}}{4\pi^2}\sum_{n=0}^\infty
(1-\frac{\delta_{n0}}{2})\int_0^\infty dp_3 e^{-(p_3^2+2\widetilde{e}\widetilde{H}n)/
\Lambda^2}[8|\varepsilon^{(c)}|+8|\overline{\varepsilon}^{(c)}|],
\end{equation}
\\

\begin{equation}
\Omega_{N} =-\frac{1}{4\pi^2}\int_0^\infty dp p^2 e^{-p^2/\Lambda^2}[6|
\varepsilon^{(0)}|+6|\overline{\varepsilon}^{(0)}|]-\frac{1}{4\pi^2}\int_0^\infty
dp p^2 e^{-p^2/\Lambda^2}\sum_{j=1}^2[2|\varepsilon_j^{(0)}|+2|\overline{\varepsilon}_j^{(0)}|]+
\frac{\Delta^2}{G}+\frac{2\Delta^2_H}{G},
\label{N}
\end{equation}
\end{widetext}
and
\begin{eqnarray*}
\varepsilon^{(c)}=\pm \sqrt{(\sqrt{p_3^2+2\widetilde{e}\widetilde{H}n}-\mu)^2+\Delta_H^2},
\\
\overline{\varepsilon}^{(c)}=\pm \sqrt{(\sqrt{p_3^2+2\widetilde{e}\widetilde{H}n}+\mu)^2+\Delta_H^2},
\end{eqnarray*}

\begin{eqnarray*}
\varepsilon^{(0)}=\pm \sqrt{(p-\mu)^2+\Delta^2},\qquad \overline{\varepsilon}^{(0)}=\pm \sqrt{(p+\mu)^2+\Delta^2},
\\
\varepsilon_1^{(0)}=\pm \sqrt{(p-\mu)^2+\Delta_a^2},\qquad \overline{\varepsilon}_1^{(0)}=\pm
\sqrt{(p+\mu)^2+\Delta_a^2},
\\
\varepsilon_2^{(0)}=\pm \sqrt{(p-\mu)^2+\Delta_b^2},\qquad \overline{\varepsilon}_2^{(0)}=\pm
\sqrt{(p+\mu)^2+\Delta_b^2},
\end{eqnarray*}
being the dispersion relations of the charged $(c)$ and neutral $(0)$ quarks. In the above we used the notation
\begin{eqnarray*}
\Delta_{a/b}^2=\frac{1}{4}(\Delta\pm\sqrt{\Delta^2+8\Delta_H^2})^2
\end{eqnarray*}
The MCFL gaps $\Delta$ and $\Delta_{H}$ correspond to the case
where the $(d,s)$ pairing gap, which takes place only between
neutral quarks, is $\Delta_{1}=\Delta$, while the $(u,s)$ and
$(u,d)$ pairing gaps, which receive contribution from pairs of
charged and neutral quarks, become
$\Delta_{2}=\Delta_{3}=\Delta_{H}$. The separation of the gap in
two different parameters in the MCFL case, as compared to the CFL,
where $\Delta_{1}=\Delta_{2}=\Delta_{3}$, reflects the symmetry
difference between these two phases \cite{Cristina-1}. Here again, $\Lambda$-dependent smooth cutoffs were introduced.

The effects of confinement can be incorporated by adding a bag constant $B$ to
both $\Omega_{CFL}$  and $\Omega_{MCFL}$. Besides, in the magnetized system the pure Maxwell
contribution, $\widetilde{H}^2/2$, should also be added \cite{H-Estimate}.
Hence, the thermodynamic potential of each phase is given by

\begin{equation}\label{Omega-0}
\Omega_{0}=\Omega_{CFL}+B,
\end{equation}
and
 \begin{equation}\label{Omega-H}
 \Omega_{H}=\Omega_{MCFL}+B+\frac{\widetilde{H}^2}{2},
\end{equation}
respectively.

While $\Lambda$ and $B$ must be given to solve the
system, the gaps $\Delta_{CFL}$, $\Delta$, and $\Delta_H$ have to
be found from their respective gap equations

\begin{equation}
\frac{\partial \Omega_{CFL}}{\partial \Delta_{CFL}}=0,
\label{Delta-CFL}
\end{equation}

\begin{equation}
\frac{\partial \Omega_{MCFL}}{\partial \Delta}=0,\qquad\qquad \frac{\partial \Omega_{MCFL}}{\partial \Delta_H}=0.
\label{Delta-MCFL}
\end{equation}

It is worth to mention that if we take into account the particle-antiparticle channels in the NJL model here considered, it is possible to claim that the bag pressure can be explicitly calculated in the chiral limit of this model as an effective bag "constant" that depends on the dynamical masses and chiral condensates. This was done in \cite{Buballa} by adopting a particular version of the NJL-model \cite{Klenvansky etal} that had four- and six-point interaction terms. At the high densities required for the realization of both the CFL and MCFL phases, the NJL-derived bag pressure contribution to the thermodynamic potential would reduce to its zero density value \cite{Buballa}. A natural question in the context of the present work is whether the external magnetic field could effectively modify the vacuum pressure found in \cite{Buballa}. It turns out that no significant modification can occur for field strengths below $10^{20}G$, as shown in the Appendix. Therefore for the range of fields relevant for our calculations, if one were to adopt the same model as in \cite{Buballa} the field effects can be ignored.

We must also underline that the fact that this NJL-derived bag constant is practically insensible to the magnetic field for a realistic range of field strengths does not prevent the "actual" bag constant in general to be significantly sensible to the magnetic field. First of all, the bag constant obtained within a NJL model is model-dependent. Besides, a well-known shortcoming of the NJL theory is that it cannot describe the confinement-deconfinement transition, which is a basic feature of QCD and the one most directly relevant for introducing a bag constant in a model of unconfined quarks. On the other hand, it should be highlighted that the CFL and MCFL phases can be also found independently of any NJL model, using weak-coupled QCD in the limit of asymptotically large densities. In this case the bag pressure cannot be explicitly calculated, so one has to rely on the MIT model analysis to impose some restrictions to the range of values it can take. Therefore, throughout the present paper we assume we have an undetermined bag pressure B which may or not depend on the magnetic field. Below, unless otherwise specified, whenever a fixed value of the bag constant is used, we take B=58 MeV/fm$^{3}$, which is compatible with both the MIT model and the zero density value of B found in \cite{Buballa}.

\section{Equations of State}

In this Section we derive the EoS for the CFL and MCFL phases using their respective
thermodynamic potentials (\ref{Omega-0}) and (\ref{Omega-H}), along with
their gap solutions obtained from (\ref{Delta-CFL}) and (\ref{Delta-MCFL}), respectively. The values of the free parameters $G$ and $\Lambda$ are
chosen to produce a CFL gap $\Delta_{CFL}=10$ MeV, which is within
the plausible range of values that $\Delta_{CFL}$ can take in
nature \cite{CSReviews}, and is small enough to decrease the
dependence of our results on the scale $\Lambda$
\cite{alf-Kouvaris-raj}. Then, throughout the entire analysis we
take: $G=4.32$ GeV$^{-2}$ and $\Lambda=1$ GeV.

As it is known \cite {Landau-Lifshitz}, the energy density and pressures can be obtained from the different components of the macroscopic energy-momentum tensor. In the reference frame comoving with the many-particle system, the system normal stresses (pressures) can be obtained from the diagonal spatial components, the system energy density, from the zeroth diagonal component, and the shear stresses (which are absent for the case of a uniform magnetic field) from the off-diagonal spatial components. Then, the energy-density, longitudinal and transverse pressures of the dense magnetized system are given respectively by
\begin{equation}\label{T-average}
\varepsilon=\frac{1}{\beta V}\langle \widetilde{\tau}^{00}\rangle,\quad
p_\parallel=\frac{1}{\beta V}\langle \widetilde{\tau}^{33}\rangle, \quad p_\bot=\frac{1}{\beta V}\langle \widetilde{\tau}^{\bot\bot}\rangle
\end{equation}
here the quantum-statistical average of the energy-momentum tensor is given by
\begin{equation}\label{T-average-1}
\langle \widetilde{\tau}^{\rho \lambda} \rangle=\frac{Tr\left[\widetilde{\tau}^{\rho\lambda}e^{-\beta(H-\mu N)}\right]}{Z}
\end{equation}
where
\begin{equation}\label{T-integral}
\widetilde{\tau}^{\rho \lambda}=\int_0^\beta d\tau \int d^3x \tau^{\rho \lambda}(\tau,x)
\end{equation}
and $Z$ is the partition function of the grand canonical ensemble given by
\begin{equation}\label{Partition-function}
Z=Tre^{\beta(H-\mu N)}
\end{equation}
with $H$ denoting the system Hamiltonian, $N$ the particle number, and $\beta$ the inverse absolute temperature.

In the CFL phase $p_\parallel=p_\bot=p$, and following the prescription (\ref{T-average}), the pressure and energy density is found as a function of the thermodynamic potential (\ref{Omega-0}) as
\begin{subequations}
\begin{equation}
\epsilon_{CFL}=\Omega_{0}-\mu \frac{\partial \Omega_{0}}{\partial \mu},
\end{equation}
\begin{equation}
p_{CFL}=-\Omega_{0}
\end{equation}\label{EoS-CFL}
\end{subequations}
while for the MCFL, due to the anisotropy introduced by the uniform magnetic field, $p_\parallel\neq p_\bot$, and the energy density and pressures are found as function of the thermodynamic potential (\ref{Omega-H}) as (see Ref. \cite{H-Estimate} for detailed derivations of the formulas for the pressures and energy density in a magnetic field)

\begin{subequations}
\begin{equation} \label{EOS-MCFL}
\epsilon_{MCFL}=\Omega_{H}-\mu \frac{\partial \Omega_{H}}{\partial \mu},
\end{equation}
\begin{equation}
p^\|_{MCFL}=-\Omega_{H},
\end{equation}
\begin{equation}
 p^\bot_{MCFL}=-\Omega_{H}+\widetilde{H} \frac{\partial\Omega_{H}}{\partial \widetilde{H}}
\end{equation}
\end{subequations}

Notice that in the MCFL phase, because of the presence of the
magnetic field, there is a splitting between the parallel $p^\|_{MCFL}$ (i.e.
along the field) and the transverse $p^\bot_{MCFL}$ (i.e.
perpendicular to the field) pressures. We call attention that
in Eq. (\ref{EoS-CFL})
(Eq. (\ref{EOS-MCFL})) the gap is a function of $\mu$
($\mu$ and $\widetilde{H}$) found by solving Eq. (\ref{Delta-CFL})
(Eq.(\ref{Delta-MCFL})). The anisotropic nature of the
system in the MCFL phase is an important feature that will be
discussed later in connection to stellar models.

\begin{figure}
\begin{center}
\includegraphics[width=0.47\textwidth]{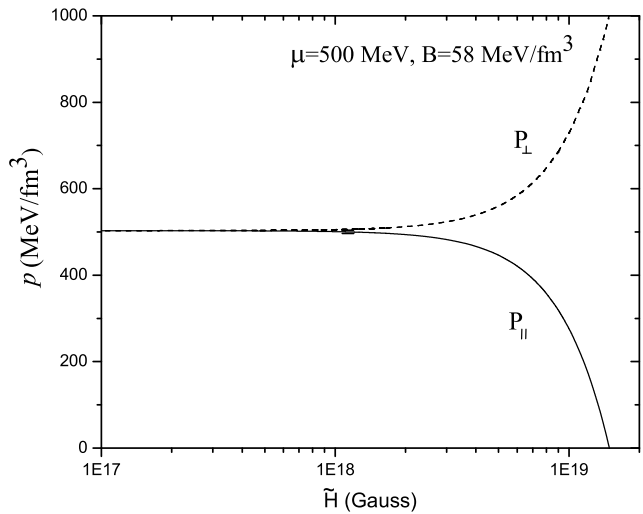}
\caption{\footnotesize Parallel and perpendicular pressures as a
function of the magnetic field intensity for representative values
of $\mu$ and bag constant $B$.} \label{Pressures}
\end{center}
\end{figure}

The magnetic field dependencies of the parallel and transverse
pressures in (\ref{EOS-MCFL}) are plotted in Fig. \ref{Pressures}. Similarly
to what occurs in the case of a magnetized uncoupled fermion system at finite
density \cite{H-Estimate}, the transverse pressure in the MCFL phase increases with the field,
while the parallel pressure decreases and reaches a zero value at
field strength of order $\gtrsim 10^{19}$ G for the density under
consideration ($\mu=500$ MeV). We see from Fig. \ref{Pressures} that $\Omega_H$ and
$\partial \Omega_H/\partial \widetilde{H}$ do not exhibit the Hass-van Alphen
oscillations as happens with other physical quantities in the presence of a magnetic
field \cite{Noronha, Fukushima, Klimenko}. This is due to the high contribution of the
pure Maxwell term in $\Omega_H$ and $\partial \Omega_H/\partial \widetilde{H}$,
which makes the oscillations of the matter part negligible in comparison.

\begin{figure}
\begin{center}
\includegraphics[width=0.44\textwidth]{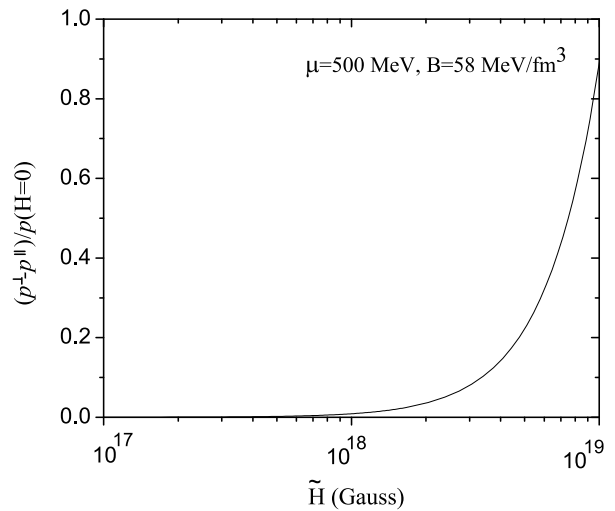}
\caption{\footnotesize Splitting of the parallel and perpendicular
pressures, normalized to the zero value pressure ($p(H=0)$), as a
function of the magnetic field intensity for $\mu=500$ MeV and
$B=58$ MeV/fm$^3$.} \label{Split}
\end{center}
\end{figure}

The splitting between parallel and perpendicular pressures, shown
in the vertical axis of Fig.\ref{Split}, grows with the magnetic
field strength. Comparing the found splitting with the pressure of the (isotropic) CFL
phase, we can address how important this effect is for the
EoS. Notice that for $3 \times 10^{18}$ G the
pressures splitting is $\sim 10\%$ of their isotropic value at
zero field (i.e. the one corresponding to the CFL phase).

\begin{figure}
\begin{center}
\includegraphics[width=0.47\textwidth]{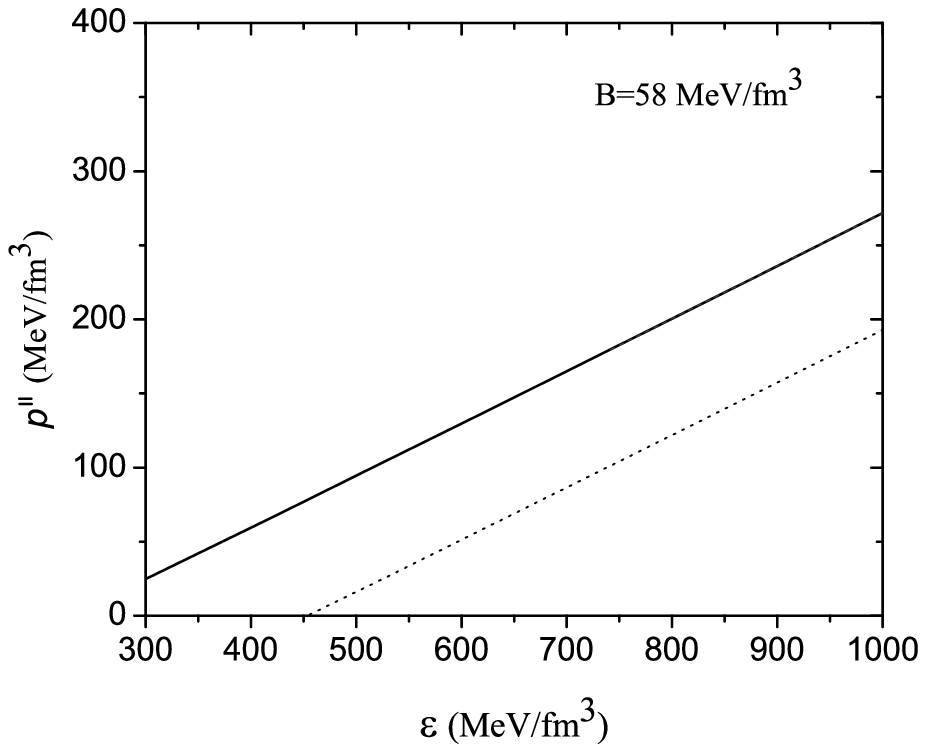}
\includegraphics[width=0.47\textwidth]{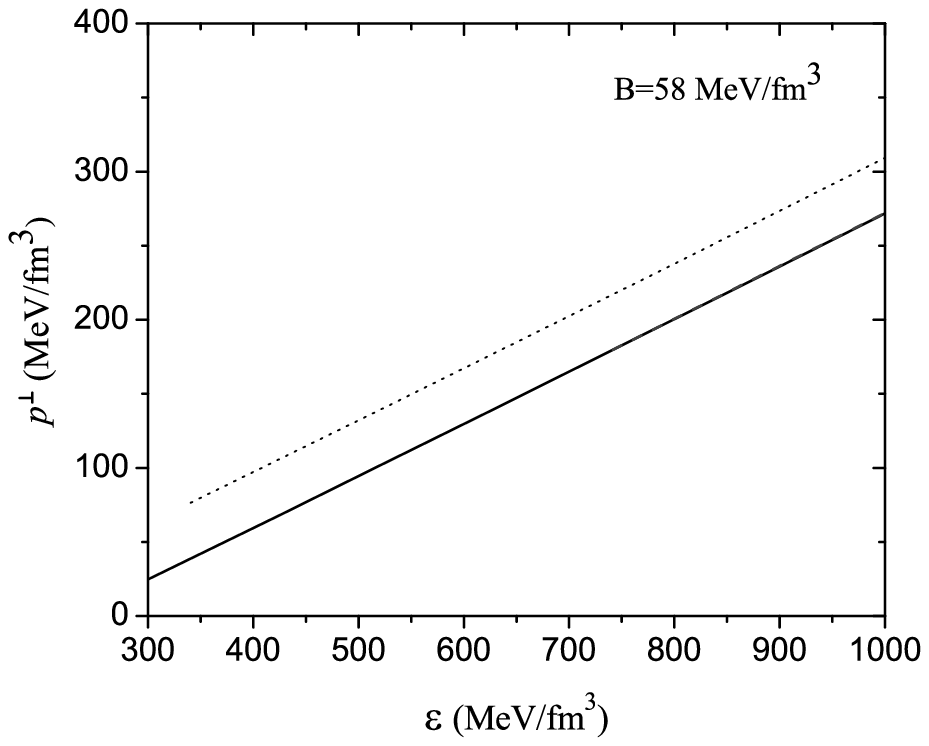}
\caption{\footnotesize Equation of state for MCFL matter
considering parallel (upper panel) and perpendicular (lower panel)
pressures for different values of $\tilde{H}$: zero field (solid line),
$10^{17} G$ (dashed line) and $5 \times 10^{18} G$ (dotted line).
Note that the low value of $H=10^{17} G$ is not distinguishable in
the plots, being merged with the zero-field curve. The value of
the bag constant was fixed to B=58 MeV/fm$^3$.} \label{EOS}
\end{center}
\end{figure}

In the graphical representation of the EoS in Fig. \ref{EOS} the
highly anisotropic behavior of the magnetized medium is explicitly
shown. While the magnetic-field effect is significant for the
$\epsilon-p^\|$ relationship at $\widetilde{H}\sim 10^{18}$ G, with a
shift in the energy density with respect to the zero-field value of
$\sim 200$ MeV/fm$^3$ for the same pressure, the field effect in the
$\epsilon-p^\bot$ relationship is smaller for the same
range of field values.

\begin{figure}
\begin{center}
\includegraphics[width=0.44\textwidth]{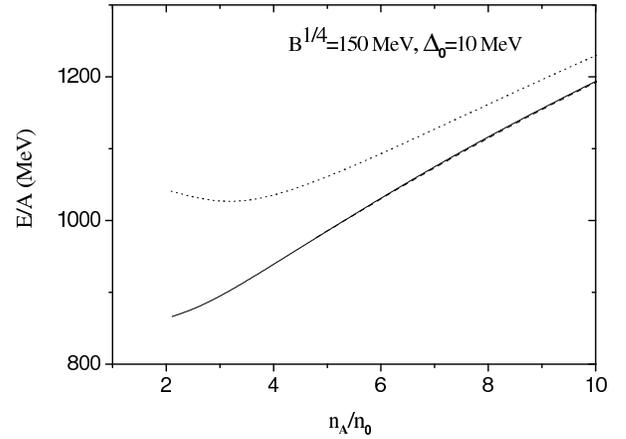}
\caption{\footnotesize Energy per baryon number as a function of
the baryonic density of MCFL matter for different values of the
magnetic field, labeled as in Fig.3. We see that increasing the magnetic field increases the energy per baryon, thus
making the matter less stable.} \label{EnergyA}
\end{center}
\end{figure}

If we use
\begin{equation}\label{mu-N}
-\frac{ \partial\Omega_{H}}{\partial\mu}=-\frac{ \partial\Omega_{MCFL}}{\partial\mu}=N,
\end{equation}
to express the chemical potential $\mu$ in terms of the baryon number density N, plug it into the gaps equations (\ref{Delta-MCFL}) to find the gaps in terms of the field $\widetilde{H}$ and N, and substitute everything into the energy density (\ref{EOS-MCFL}), we can see that the energy density per baryon number (Fig. \ref{EnergyA}) {\it increases} with increasing magnetic
field, in contrast to previous claims based on a CFL model at
$\widetilde{H}\neq 0$ with only one gap that was fixed by hand
\cite{Aurora}.

\section{Stability conditions}

Having the EoS for MCFL matter, we can analyze the conditions for
this matter to become absolutely stable. This is done by comparing the
energy density at zero pressure condition with that of the iron nucleus (roughly 930 MeV). Depending on whether the energy density of the MCFL phase is higher or smaller than this value, the content of a magnetized strange quark could be or not made of MCFL matter. If
the energy of the MCFL phase is smaller than 930 MeV for only a specific range in
pressure (or density), this would imply metastability.

To find the maximum value of the bag constant required for the stability to hold at zero magnetic field, and
then use it as a reference when considering the MCFL case, we will start our analysis investigating the stability
in the CFL phase. We call reader's attention that in all our derivations we work within a self-consistent approach, in which the solutions of the gap equations are substituted in the pressures and energies of each phase before imposing the conditions of equilibrium and stability.

\subsection{$H=0$ Case}

The stability criterion for CFL matter in the absence of a magnetic
field is very simple. Following Farhi and Jaffe's \cite{Farhi84} approach, we can
determine the maximum value of the bag constant that satisfies
the stability condition at zero pressure. With this aim, we first impose
the zero pressure condition in Eq. (\ref{EoS-CFL}), to get
\begin{eqnarray}  \label{CFL-Equilibrium}
B&=&-\Omega_{CFL} \\
\epsilon_{CFL}&=&-\mu \frac{\partial \Omega_{CFL}}{\partial \mu}
\end{eqnarray}

Taking into account that in the CFL phase each of the three flavors have
the same number density (which is correct as long as one does not
introduce the strange mass and has to impose charge neutrality),
we have $n_A=\frac{1}{3}(n_u+n_d+n_s)=\frac{1}{3}N$. Hence, the
energy density per particle becomes

\begin{equation}  \label{Energy-0}
\frac{\epsilon_{CFL}}{n_A}=-\frac{\mu_0}{n_A} \frac{\partial \Omega_{CFL}}
{\partial \mu}|_{\mu_0}=\frac{\mu_0}{n_A}N|_{\mu_0}=3\mu_0
\end{equation}
with $\mu_0$ denoting the chemical potential at zero pressure.
For the CFL matter to be absolutely stable, its energy density per particle should be smaller than the lowest energy density per baryon found in nuclei, i.e. that corresponding to the iron nucleus. Hence,
\begin{equation}  \label{Energy-nuclear}
\frac{\epsilon_{CFL}}{n_A} \leq \epsilon_{0}(Fe^{56}),
\end{equation}
where $\epsilon_{0}(Fe^{56}) =\frac{1}{56} m(Fe^{56})\approx 930 MeV$. This condition constraints the maximum allowed value of the chemical potential to be $\mu_0=310
 MeV$. Using this result back in (\ref{CFL-Equilibrium}) we can determine the value of the
maximum bag constant for absolute stability to hold. The obtained result is shown in Fig.
\ref{window} (horizontal axis). This bag constant value is within an acceptable range.
Moreover, it is of the same order as the one given
in reference \cite{German} for $m_s=0$.

\subsection{$H\neq 0$ Case}

When $H\neq 0$ the situation is different. Now, both the parallel
and perpendicular pressures in Eq. (\ref{EOS-MCFL}) need to vanish
simultaneously. Therefore, the two equilibrium conditions become

\begin{eqnarray}
p^\|_{MCFL}&=&-\Omega_{MCFL}-B-\frac{\widetilde{H}^2}{2}=0, \label{MCFL-Equilibrium-1}
\\
 p^\bot_{MCFL}&=&\widetilde{H} \frac{\partial\Omega_{MCFL}}{\partial \widetilde{H}}+\widetilde{H}
 \frac{\partial B}{\partial \widetilde{H}}+\widetilde{H}^2=0  \label{MCFL-Equilibrium-2}
\end{eqnarray}

Where we are assuming that the bag constant depends on the magnetic
field. It is not unnatural to expect that the applied magnetic
field could modify the QCD vacuum, hence producing a field-dependent bag constant. One can readily
verify that Eqs.
(\ref{MCFL-Equilibrium-1})-(\ref{MCFL-Equilibrium-2}) are
equivalent to require $p^\|_{MCFL}=0$ and $\partial
p^\|_{MCFL}/\partial \widetilde{H}=0$ at the equilibrium point.

Equation (\ref{MCFL-Equilibrium-2}) can be rewritten as
\begin{equation}  \label{Equilibrium-3}
\widetilde{H}=M-\frac{\partial B}{\partial \widetilde{H}}
\end{equation}
where $M=-\partial \Omega_{MCFL}/\partial \widetilde{H}$ is the
system magnetization. If we were to consider that the vacuum energy
$B$ does not depend on the magnetic field, we would need
\begin{equation}  \label{Magnetization}
M=\widetilde{H},
\end{equation}
to ensure the equilibrium of the self-bound matter, a condition difficult to satisfy since
it would imply that the medium response to the applied magnetic
field (i.e. the medium magnetization $M$) is of the order of the
applied field that produces it. Only if the MCFL matter were a ferromagnet this would be viable. The other possibility for the
equilibrium conditions (\ref{MCFL-Equilibrium-1}) and
(\ref{MCFL-Equilibrium-2}) to hold simultaneously is to have a
field-dependent bag constant capable to yield nonzero vacuum magnetization
$M_0=-\frac{\partial B}{\partial \widetilde{H}}\simeq \widetilde{H}$. From the discussion at the end of Section II and the results of the Appendix, it is clear that this condition cannot be satisfied if the bag constant were the one found in \cite{Buballa}. However, as argued before, such a bag constant is model-dependent and was obtained within a theory that does not exhibit confinement.  Hence, we cannot discard the possibility that the actual bag constant is much more sensitive to the applied magnetic field. We must recall that in
other physical scenarios, bag constants depending on external
conditions such as temperature
and/or density have been previously considered
\cite{mu-T-dependent-B}. Luckily, in the approach we follow here we do not need to
formulate a theory for the $\widetilde{H}$-modified vacuum, as we only need to know the relation between $B$ and $\widetilde{H}$ under equilibrium conditions.

The following comment is in order. The fact that the bag constant
needs to be field-dependent for self-bound stars in a strong magnetic field is a direct consequence of the lack of
a compensating effect for the internal pressure produced by the
magnetic field other than that applied by the vacuum (an exception could be of course if the paired quark matter would exhibit ferromagnetism). For gravitationally bound stars,
on the other hand, the situation is different, since the own
gravitational field can supply the pressure to compensate the one
due to the field. For such systems, keeping $B$ constant in
the EoS is in principle possible. Under this assumption we
considered a fixed $B$-value in Fig. \ref{EOS}.

To determine the maximum "bag constant" allowed for each magnetic field
value in the stable region, we need to simultaneously satisfy the equilibrium equations (\ref{MCFL-Equilibrium-1}) and
(\ref{MCFL-Equilibrium-2}), as well as the stability condition in the presence of the magnetic field
\begin{eqnarray}  \label{Energy-P0}
\frac{\epsilon_{MCFL}}{n_A}=-\frac{\mu_{\widetilde{H}}}{n_A} \frac{\partial
\Omega_{MCFL}}{\partial \mu}|_{\mu_{\widetilde{H}}}-\frac{\widetilde{H}^{2}}{2n_A}\nonumber
\\
=3\mu_{\widetilde{H}}-3\frac{\widetilde{H}^{2}}{2N}\leq \epsilon_{\widetilde{H}}(Fe^{56})
\end{eqnarray}
Notice that because the nucleons' rest-energy are modified in the presence of a magnetic field, the energy density of iron $\epsilon_{\widetilde{H}}(Fe^{56})$ is now field-dependent. Taking into account the field interaction with the anomalous magnetic moment \cite{Broderick}, the nucleons' energy spectrum at $H\neq 0$, is given by
\begin{equation}\label{nucleon-energy}
E_i=\sqrt{[\sqrt{c^4m_i^2+c^2(p_\bot^2)_i}+\kappa_iH\sigma]^2+c^2p_z^2},\qquad i=p,n
\end{equation}
For the proton ($i=p$), and neutron ($i=n$), the following parameters hold respectively,
\begin{eqnarray}\label{proton}
m_p=938.28 MeV,\quad \kappa_p=\mu_N(g_p/2-1),\nonumber
\\
(p_\bot^2)_p=2leH,\quad l=0,1,2,..., \qquad
\end{eqnarray}
\begin{eqnarray}\label{neutron}
m_n=939.57 MeV,\quad \kappa_n=\mu_Ng_n/2,\nonumber
\\
 (p_\bot^2)_n=p_1^2+p_2^2,\qquad\qquad
\end{eqnarray}
In (\ref{proton})-(\ref{neutron}), $\mu_N=e\hbar /2cm_p$ is the nuclear magneton, and the Lande g factors are given by $g_p=5.58$ and $g_n=-3.82$, respectively.

The proton and neutron rest energies can be obtained from (\ref{nucleon-energy}) at zero momentum
\begin{eqnarray}\label{nucleon-rest-energy}
E_p^{(0)}=m_pc^2+\frac{\sigma}{2}(g_p/2-1)\frac{e\hbar H}{m_pc},\nonumber
\\
 E_n^{(0)}=m_nc^2+\frac{\sigma}{2}g_n/2\frac{e\hbar H}{m_nc}\quad\qquad
\end{eqnarray}

It would take a magnetic-field strength larger than $10^{20} G$ to have the second terms in the RHS of Eqs. (\ref{nucleon-rest-energy}) comparable to the first ones. For the field range considered in this paper ($H\leq 10^{19} G$) it is then consistent to neglect the field correction in the iron energy density, thus making $\epsilon_{\widetilde{H}}(Fe^{56})\approx \epsilon_{0}(Fe^{56}) =930 MeV$.

\begin{figure}
\begin{center}
\includegraphics[width=0.44\textwidth]{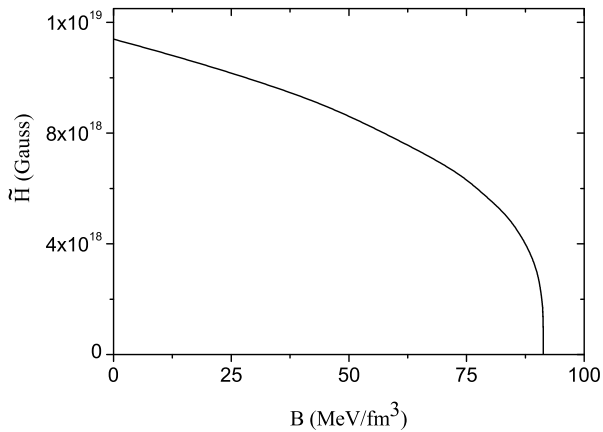}
\caption{\footnotesize Stability window for MCFL matter in the
plane $\tilde{H}$ vs. $B$. The curve shown corresponds to the
borderline value $\epsilon/A=930$ MeV.} \label{window}
\end{center}
\end{figure}

Then, finding $\mu_{\widetilde{H}}$ as a function of $\widetilde{H}$ in (\ref{Energy-P0}) and substituting it back in (\ref{MCFL-Equilibrium-1}), we can numerically solve
\begin{equation}\label{bagH}
B(\widetilde{H})=-\Omega_{MCFL}(\mu_{\widetilde{H}},\widetilde{H})-\widetilde{H}^2/2,
\end{equation}
to determine the stability window in the plane
$\widetilde{H}$ versus $B$ for the MCFL matter to be absolute stable
(Fig. \ref{window}). The
inner region, which corresponds to smaller bag constants for each given
$\widetilde{H}$, is the absolutely stable region.

Note that, contrary to Farhi and Jaffe \cite{Farhi84}, we did not impose a {\it minimum} value for the bag
constant because we have no clear indication from experiments of the
possible behavior of this parameter when a magnetic field is
applied to a system.

\begin{figure}
\begin{center}
\includegraphics[width=0.44\textwidth]{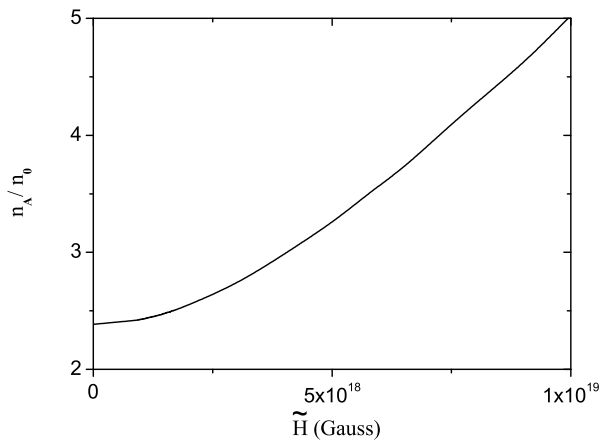}
\caption{\footnotesize Baryonic density at zero pressure conditions
for MCFL matter as a function of the magnetic field $\widetilde{H}$
considering the field dependence of the bag constant given in Eq.
(\ref{bagH}).} \label{density}
\end{center}
\end{figure}

As shown in Fig.
\ref{density}, the value of the chemical potential $\mu_{\widetilde{H}}$ found from
the stability condition (\ref{Energy-P0}), grows with increasing
${\widetilde{H}}$, in good consistency with our assumption of zero quark masses and deconfined quark matter. In summary, our results indicate that a condition for the MCFL matter to be absolutely stable is to have a field-dependent bag constant.

\section{Stellar models}

The most immediate application of the EoS for the MCFL phase
is to construct stellar models for compact stars composed of quark matter.
There are two distinct possibilities: new magnetized ``strange stars'', if quark matter
in the MCFL phase is absolute
stable (the possibility explored in the last section); and hybrid stars, if the
MCFL matter is metastable (stars would contain a MCFL core surrounded by normal
matter).

As long as the magnetic field strength is not much larger than the threshold value $\sim 10^{18} G$, at which the pressure anisotropy starts to become noticeable, both cases can be investigated by integrating the relativistic
equations for stellar structure, that is, the
Tolman-Oppenheimer-Volkoff (TOV) and mass continuity equations,
\begin{eqnarray}
\frac{dm}{dr}&=&4\pi r^2\epsilon \label{TOV1}\\
\frac{dP}{dr}&=&-\frac{\epsilon m}{r^2}\Big(1+\frac{P}{\epsilon}\Big)\Big
(1+\frac{4\pi r^3P}{m}\Big)\Big(1-\frac{2m}{r}\Big)^{-1}\label{TOV2}
\end{eqnarray}
written in natural units, $c = G = 1$. Given that this set of differential equations apply only to isotropic EoS, while our
results for the pressures indicate a rapidly growing anisotropy of the EoS beyond the threshold field (Fig.1),
our approach is probing the limits of the validity of spherical models based on
isotropic EoS.

\subsection{Magnetic CFL Strange Stars}

\begin{figure}
\begin{center}
\includegraphics[width=0.5\textwidth]{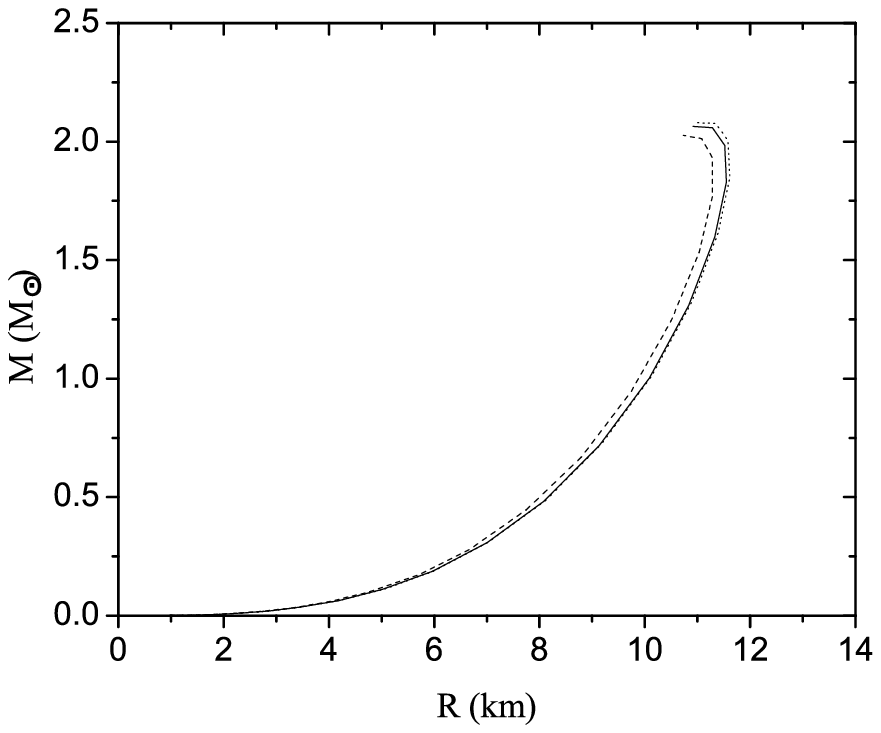}
\includegraphics[width=0.5\textwidth]{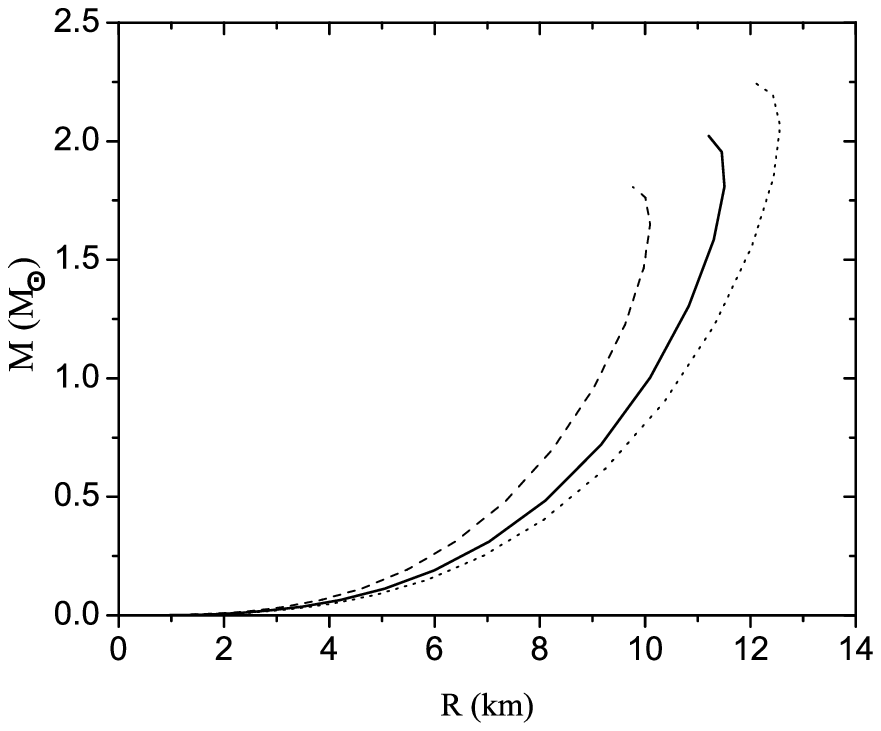}
\caption{\footnotesize Mass-radius relation for magnetized strange
CFL stars and bag constant $B=58$ MeV/fm$^3$. The full line
indicates the M-R relation for zero magnetic field, whereas the
dashed and dotted lines represent the MR relation calculated with
the parallel and perpendicular pressures, respectively, for
$\tilde{H}=1.7\times 10^{17}$ G (upper panel) and
$\tilde{H}=3\times 10^{18}$ G (lower panel).} \label{MR}
\end{center}
\end{figure}

Based on the analogy with Ref. \cite{AFO, Witten}, we construct stellar models using the
EoS with parameters inside the stability window, that is, for a self-bound matter case.
In Fig. \ref{MR} we present the mass-radius relation for two values
of the magnetic field, when the anisotropy is still small (a few parts per thousand, see
Fig. \ref{EOS}) and when the anisotropy cannot be neglected (a few per cent, Fig. 3).
For each of these values of the field we have calculated two curves, one
considering the pressure given by the parallel (dotted line) and the other
given by the perpendicular one (dashed line), and compared them with the zero
field mass-radius relation in Fig. \ref{MR}.

\begin{figure}
\begin{center}
\includegraphics[width=0.5\textwidth]{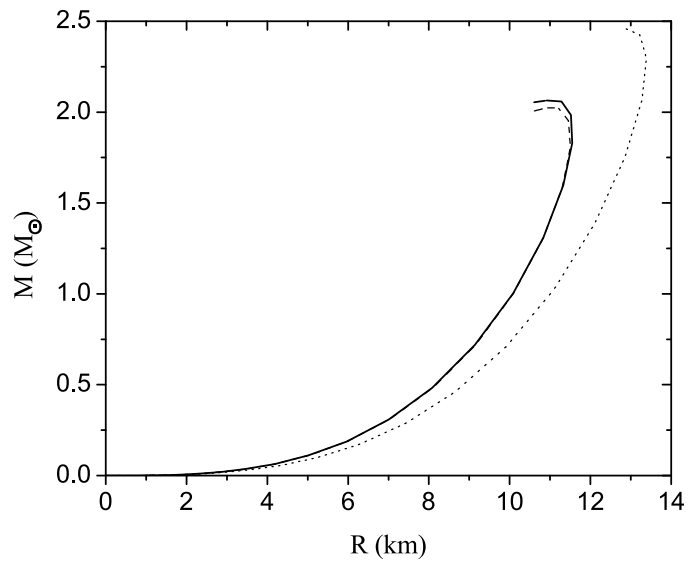}
\caption{\footnotesize Mass-radius relation for the EoS given in
\cite{German} for CFL matter without magnetic field for two
different values of the gap parameter, $\Delta =0$ (dashed) and
$\Delta= 100 MeV$, and the results obtained here setting $\tilde H
=0$ (solid line).} \label{Comp}
\end{center}
\end{figure}

Even though the calculations in Fig. \ref{MR} should be considered
as just an example, we see that the perpendicular pressure
provides a ``harder EOS'' while the parallel is ``softer''.
Therefore, the former choice renders a higher maximum mass.

From those Figs. we conclude that one must restrict oneself to
weak magnetic fields, when the deviation from spherical symmetry
is very small (of order 0.001 \%), in order to justify the use of
Eqs. (\ref{TOV1}) and (\ref{TOV2}). If the magnetic field in
these compact stars is too high, say $\tilde{H} \gtrsim 10^{18}$
G  (at $\mu=500$ MeV), the spherically symmetric TOV equations cannot be employed
because the deviations become important and lead to significant
differences with respect of realistic axisymmetric models, yet to
be constructed taking into account the pressure asymmetry. (this is why the M-R sequences in Fig. \ref{MR}
lower panel should not be trusted, and we stress again that they
should be considered just as an example). We shall address this
issue elsewhere.

Fig. \ref{Comp} compares our result for zero magnetic field EoS
with the one presented in reference \cite{German}. It shows
that the curves calculated using NJL + B and bag model with
small $\Delta$ are quite similar. Nonetheless, it is important
to stress that the calculations of the present work feature a self-consistent gap parameter
(not a constant), which varies according to the particle density.

\subsection{Hybrid Stars}

The construction of models for the so-called ``hybrid stars''
faces the same problem as before when the magnetic-field-induced
pressure anisotropy is considered. Working outside the
stability windows render EoS which are valid only above a certain
critical density, not all the way down to zero, since MCFL matter
would be favored at high density only. Thus, the stellar models
belong to the so-called {\it hybrid} type, in which a core of the
high-density matter is present. Again, the value of the magnetic
field induces an increasingly large anisotropy, which in turn
forces the construction of axisymmetric (not spherical) stellar
models. In this way, it can be modelled within the isotropic TOV
formalism only for relative pressure differences in the ballpark
of $\sim 10^{-3}$.

\begin{figure}
\begin{center}
\includegraphics[width=0.5\textwidth]{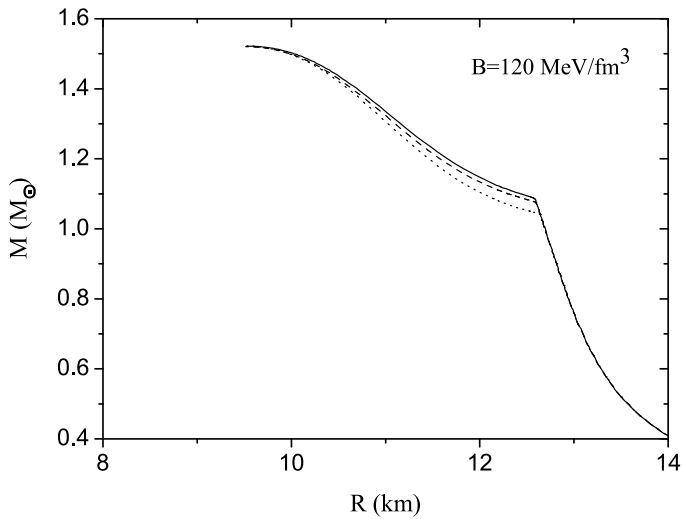}
\caption{\footnotesize Mass-radius relation for hybrid stars. The
inner core is composed of CFL matter with the corresponding
magnetic field. From the transition density to the surface of the
star, we have employed the zero-temperature Bethe-Johnson EoS. The
stellar sequences correspond to zero (solid), $2 \times 10^{17} G$
(dashed) and $10^{18} G$ (dotted) fields respectively.}
\label{Hybrid}
\end{center}
\end{figure}

Fig. \ref{Hybrid} displays a hybrid sequence obtained by employing
the well-known Bethe-Johnson EoS for nuclear matter and using the
Gibbs criteria for determining the value of the transition
pressure between exotic and nuclear matter. These curves were
calculated using the perpendicular pressure (for magnetized stars)
as an example. As expected, the appearance of an MCFL core
{\it softens} the EoS, rendering lower maximum masses than
``pure'' hadronic models. The main feature of considering the
existence of magnetic field for hybrid MCFL stars is to switch the
point where the hybrid sequences begin, i. e. where the stars
start exhibiting a CFL core. Since the difference in the EoS for
low field MCFL matter and CFL matter is of just a few percent (see
Fig. \ref{EOS}), and because the star radius depends mainly on the nuclear
EoS, observational data of maximum mass and minimum radii would
not be able to distinguish the existence of low magnetic fields in
these hybrid stars. Again, for high magnetic fields
($\tilde{H}\gtrsim 10^{18}$ G) the results are still to be
analyzed, but differences in the maximum allowed mass may arise,
and therefore the results of spherical models cannot be trusted.
This is potentially important for the identification of actual
compact stars masses and radii \cite{Ozels} (see next
Section).

\section{Conclusions}

We have shown that a magnetic field in CFL matter (termed MCFL
phase here) does not, generally speaking, favor the stability scenario, and
even forces a new condition (a field dependent vacuum ``bag
constant'') which is perhaps physically reasonable but cannot be
verified as yet. However, absolute stability is not excluded provided
the vacuum is properly modified by the magnetic field. On the
other hand, in the absence of this dependence, we conclude that
there is no room for absolute stability of CFL matter under the
influence of a magnetic field within the model. If this is the
case, there could be no magnetized ``strange stars'', but only
hybrid stars. Even before calculating stellar models in the
anisotropic pressure regime, we can state that the found stability
conditions can impose a maximum magnetic field that could be
supported by self-bound MCFL strange stars (that is, stars made of
this magnetized self-bound paired matter), a feature which in
principle, could be compared with observations.

In the self-consistent approach used here for the gap parameter, we
do not find much difference in the stability region at zero field as
compared to the case in which the gap parameter is parameterized (and
extended to quite high values) \cite{German}. The EoS
is still largely linear and substantially modified only at
sufficiently high fields where the magnetized medium becomes highly
anisotropic. It is not surprising then that in the quasi-isotropic regime
($H \leq 10^{18}$ G) the stellar sequences are not very different from the
zero-field case (see Fig. \ref{MR}).

We should notice that the anisotropic pressure regime is
attainable at field values that are allowed in the core of compact
stellar objects \cite{H-Estimate}. From the heuristic arguments presented
in the Introduction, and then analytically and numerically corroborated in the paper,
field effects become relevant in the EoS of MCFL matter for field strength
$\widetilde{H}\gtrsim 10^{18}G$. Nevertheless, for that field range the system
asymmetry, expressed in terms of the pressure splitting
$(\Delta p/p_{CFL})\sim (\widetilde{H}^2/\mu^2\Delta^2)\sim {\cal{O}}(1)$, is significant,
then invalidating the use of the TOV formalism. Thus, to work in the anisotropic
regime, where the most interesting field effects should occur,
an entirely different stellar structure formalism in agreement
with the system cylindrical symmetry would be needed,
since the spherical symmetry is broken from the very beginning by the
presence of the strong magnetic field.
We underline that the conventional TOV equations were obtained by solving the
Einstein equations for a
general time-invariant, spherically symmetric metric. That is, they were
derived assuming spherically symmetric and isotropic
medium in static gravitational equilibrium. Hence, it becomes
imperative to generalize the TOV equations to an anisotropic
medium employing a metric with cylindrical symmetry
\cite{cilindrico} that can accommodate the splitting of the
longitudinal and transverse pressures appearing at strong enough
magnetic fields. We expect to address this issue in a
future publication.

It is important to stress that recent works \cite{Ozels}-\cite{Ozels2} have
exploited the increasing availability of high-quality data to
pindown masses and radii of selected stellar systems. Even though
the results and analysis are far from being definitive, there is
evidence favoring a relatively soft EoS to model 4U 1608-52, 4U
1820-30 and EXO 1745-248 \cite{Ozels2}, at least in the region immediately
above the saturation density. Comparing the prediction of our hybrid EoS model shown in Fig. 9
with the 1- and 2-$\sigma$ confidence contours for the masses
and radii of the three neutron stars in these binary systems, (shown in Fig. 1 of Ref. \cite{Ozels2}),
one can easily see that for fields within the isotropic regime our EoS is compatible with these observations.

On the other hand, recent measurements \cite{Demorest-Nature2010} of the Shapiro delay in the
radio pulsar PSRJ1614-2230 have yielded a
mass of $1.97\pm 0.04M_{\odot}$. Even though such a large mass calls for a stiffer EoS, it does not
rule out color superconductivity in the star's core or a self-bound model like the curves in Fig. 7.
In fact, by using the phenomenological EoS for quark matter proposed in Ref. \cite{Alford EoS}, the
authors of Ref. \cite{Ozels3} showed that a large value of the star mass is only compatible with
strongly interacting quarks paired in a color superconducting state. It is an interesting open question
to explore, within the self-consistent approach used in our calculations, whether there is a physically
viable region of the parameter space of the MCFL phase that can produce EoS curves compatible
with the PSRJ1614-2230 mass observation.

Even if the measured systems do {\it not} possess a
noticeable magnetic field, future determinations of SGR-AXN radii
and masses are foreseeable. For those systems, an additional
complication would arise with the consideration of the magnetic
field, as discussed above. The particular case of MCFL elaborated
here suggests that a full evaluation that takes into account the pressure anisotropy may be
necessary to address masses and radii in the presence of very strong magnetic fields.

\begin{acknowledgments}
L.P. and J.E.H. acknowledge the financial support received from
the Funda\c c\~ao de Amparo \`a Pesquisa do Estado de S\~ao Paulo.
They also thank UTEP for support and hospitality during three scientific visits
in which this work was performed.
J.E.H. also wishes to acknowledge the CNPq Agency (Brazil) for
partial financial support. We thank Marcio G. B. de Avellar for
useful discussions on mass-radius relation calculations. E.J.F. and
V.I like to thank the Instituto de Astronomia, Geof\'\i sica e Ci\^encias
Atmosf\'ericas,  S\~ao Paulo, for the warm hospitality extended to them during their visit.
The work of E.J.F. and V.I was supported in part by the Office of Nuclear Theory of
the Department of Energy under contract DE-FG02-09ER41599.

\end{acknowledgments}

\appendix

\section{Dynamical bag constant in the chiral limit at $H\neq 0$}
\label{Dyn-Bag-Const}

Let us investigate the effect of an external magnetic field on the bag pressure found in Ref. \cite{Buballa}. In the dynamical approach of Ref. \cite{Buballa} the bag pressure has its origin in the spontaneous breaking of chiral symmetry. For our high density system the vacuum pressure contribution found in \cite{Buballa} reduces to $B_0=B|_{n_u=n_d=n_s=0}$ taken in the chiral limit $m_{i0}=0$ with
\begin{widetext}
\begin{eqnarray}\label{Dynamical-B}
B=\sum_{i=u,d,s}\left[\frac{3}{\pi^2}\int_0^\Lambda
p^2dp\left(\sqrt{m_i^2+p^2}-\sqrt{m_{i0}^{2}+p^2}\right)
-2G \langle \overline{q}_i q_i \rangle \right]+4K\langle \overline{u}u\rangle \langle \overline{d}d\rangle \langle \overline{s}s\rangle,
\end{eqnarray}
\end{widetext}
Here  $m_{i0}$  and $m_i$ are the current and dynamical quark masses respectively, $G$ and $K$ are quark self-interacting constants with dimensions $energy^{-2}$ and $energy^{-5}$, respectively, and $\langle\overline{q}_i q_i \rangle$ are the quark condensates given as functions of the corresponding quark dynamical masses by
\begin{equation}
\langle\overline{q}_i q_i \rangle=-\frac{3}{\pi^2}\int_{p_{Fi}}^\Lambda p^2 dp \frac{m_i}{\sqrt{m_i^2+p^2}}
\label{Quark-Condensates}
\end{equation}
with $p_{Fi}=(\pi^2n_i)^{1/3}$ being the Fermi momenta depending on the densities $n_i=\langle q_i^\dag q_i\rangle$.

A magnetic field modifies the expressions for the bag $B$ and the chiral condensates in the following way
\begin{widetext}
\begin{eqnarray}\label{Dynamical-B-H}
B_H=\sum_{i=u,d,s}\left[\frac{3q_iH}{2\pi^2}\sum_{n=0}^{[\Lambda^2/q_iH]}d(n)\int_0^\Lambda
dp_3\left(\sqrt{m_i^2+\overline{p}^2}-\sqrt{m_{i0}^{2}+\overline{p}^2}\right)-2G \langle \overline{q}_i q_i \rangle_H \right]
+4K\langle \overline{u}u\rangle_H \langle \overline{d}d\rangle_H \langle \overline{s}s\rangle_H,
\end{eqnarray}
\end{widetext}
and
\begin{equation}
\langle\overline{q}_i q_i \rangle_H=-\frac{3q_iH}{2\pi^2}\sum_{n=0}^{[\Lambda^2/q_iH]}d(n)\int_{p_{Fi}}^\Lambda  dp_3 \frac{m_i}{\sqrt{m_i^2+\overline{p}^2}}
\label{Quark-Condensates-H}
\end{equation}
where we assumed a magnetic field along the $x_3$-direction, and used the notation: $\overline{p}^2=p_3^2+2q_inH$ for the three-momentum, with $n$ labeling the discrete Landau levels, $n=0,1,2,...$; $d(n)=2-\delta_{n0}$ for the spin degeneracy of the $n$ Landau level; $q_i$ for the corresponding quarks' electric charges; and $[...]$ for the integer part of the argument.

Comparing the leading term of (\ref{Dynamical-B}),
\begin{eqnarray}
B\simeq \sum_{i=u,d,s}\left[\frac{3\Lambda^2}{4\pi^2}(m_i^2-m_0^{i2})-2G \langle \overline{q}_i q_i \rangle \right]\nonumber
\\
+4K\langle \overline{u}u\rangle \langle \overline{d}d\rangle \langle \overline{s}s\rangle,\qquad\qquad\qquad\qquad
\label{Leading-B}
\end{eqnarray}
with that of (\ref{Dynamical-B-H})
\begin{eqnarray}
B_H\simeq \sum_{i=u,d,s}\left[\frac{3q_iH}{4\pi^2}[\frac{\Lambda^2}{q_iH}](m_i^2-m_0^{i2})-2G \langle \overline{q}_i q_i \rangle_H \right]\nonumber
\\
+4K\langle \overline{u}u\rangle_H \langle \overline{d}d\rangle_H \langle \overline{s}s\rangle_H,\qquad\qquad\qquad\qquad\qquad\quad
\label{Leading-B-H}
\end{eqnarray}
and taking into account that $q_iH[\frac{\Lambda^2}{q_iH}]\simeq \Lambda^2$ we have that the difference between $B$ and $B_H$ is basically due to the difference between the dynamical masses and condensates at zero and non-zero fields. That a magnetic field modifies the dynamical mass is a well known result in the literature \cite{MC}. However, as demonstrated in the NJL model for quark matter \cite{MC-NJL}, and in the QCD chiral effective theory \cite{MC-Ch-Th}, the field-induced change in the dynamical masses and the chiral condensates are negligibly small for any field smaller than $10^{20} G$. This in turns translates into a negligible modification of the vacuum pressure $B_{0}$ by magnetic fields below $10^{20} G$. Hence, for the range of fields of interest for the astrophysics of compact stars, no significant field-induced variation of the vacuum pressure found in the framework of this approach will occur.

\end{document}